\documentclass[utf8]{frontiersSCNS} 
\usepackage{url,hyperref,lineno,microtype,subcaption}
\usepackage[onehalfspacing]{setspace} 

\usepackage[dvipsnames]{xcolor}

\def\keyFont{\fontsize{8}{11}\helveticabold}
\def\firstAuthorLast{Zurbriggen {et~al.}}
\def\Authors{Ernesto Zurbriggen\,$^{1}$, C. Guillermo Gim\'enez de Castro\,$^{1,2}$, Andrea Costa\,$^{3}$, Mariana C\'ecere\,$^{3,4}$, Caius L. Selhorst\,$^{5}$}

\def\Address{$^{1}$Universidade Presbiteriana Mackenzie, Centro de R\'adio Astronomia e Astrof\'isica Mackenzie (CRAAM), SP, S\~ao Paulo, Brasil.\\
$^{2}$Consejo Nacional de Investigaciones Cient\'ificas y T\'ecnicas (CONICET), Instituto de Astronom\'ia y F\'isica del Espacio (IAFE-UBA), Buenos Aires, Argentina.\\
$^{3}$Consejo Nacional de Investigaciones Cient\'ificas y T\'ecnicas, Instituto de Astronom\'ia Te\'orica y Experimental (IATE), C\'ordoba, Argentina.\\
$^{4}$Universidad Nacional de C\'ordoba (UNC), Observatorio Astron\'omico de C\'ordoba (OAC), C\'ordoba, Argentina.\\
$^{5}$NAT - N\'ucleo de Astrof\'isica, Universidade Cidade de S\~ao Paulo, S\~ao Paulo, SP, Brasil.
}

\def\corrAuthor{Ernesto Zurbriggen}
\def\corrEmail{ernesto.zurbriggen@craam.mackenzie.br}

\begin{document} 
\onecolumn
\firstpage{1}

\title[Estimating SADs radio emission]{Estimating the coronal supra-arcade downflows radio emission: from centimetre through submillimetre wavelengths}

\author[\firstAuthorLast ]{\Authors} 
\address{} 
\correspondance{} 
\extraAuth{}

\maketitle


\begin{abstract}
\section{}
{Supra-arcade downflows (SADs) are infrequent, wiggly opaque structures observed to descend through the solar corona, mostly in EUV and soft X-ray frequencies. From their physical characteristics, SADs have been interpreted as voided (subdense) bubbles and are related to magnetic reconnection processes during long-term  erupting flares. In this work we use numerical MHD simulations to compute flux density maps, which are convolved with telescope beams to synthesise images with the aim to assess the expected SADs emission at radio wavelengths  and propose observing strategies, including the instruments that can be used. We assume that the emission is thermal bremsstrahlung from a fully ionised plasma without any appreciable gyroresonance contribution since magnetic fields are of the order of $\sim$10~G. We find that SADs emission should be optically thin in the frequency $[10$\,--\,$1000]$~GHz range, and the spatially integrated flux should be larger than $1$~Jy. We conclude, therefore, that observing SADs in radio frequencies between $[0.5$\,--\,$1000]$~GHz is feasible with present instrumentation. Moreover, since the emission is for the most part optically thin, the flux density is proportional to temperature, density and line-of-sight depth, and when combined with EUV and soft X-ray images, may allow a better density and temperature determination of SADs.}

\tiny
 \keyFont{ \section{Keywords:} Sun: corona, Sun: flares, Sun: magnetic fields, Sun: radio radiation} 
\end{abstract}

\section{Introduction} 
\label{sec:intro}      

\begin{figure}
\begin{center} 
\includegraphics[width=0.49\textwidth]{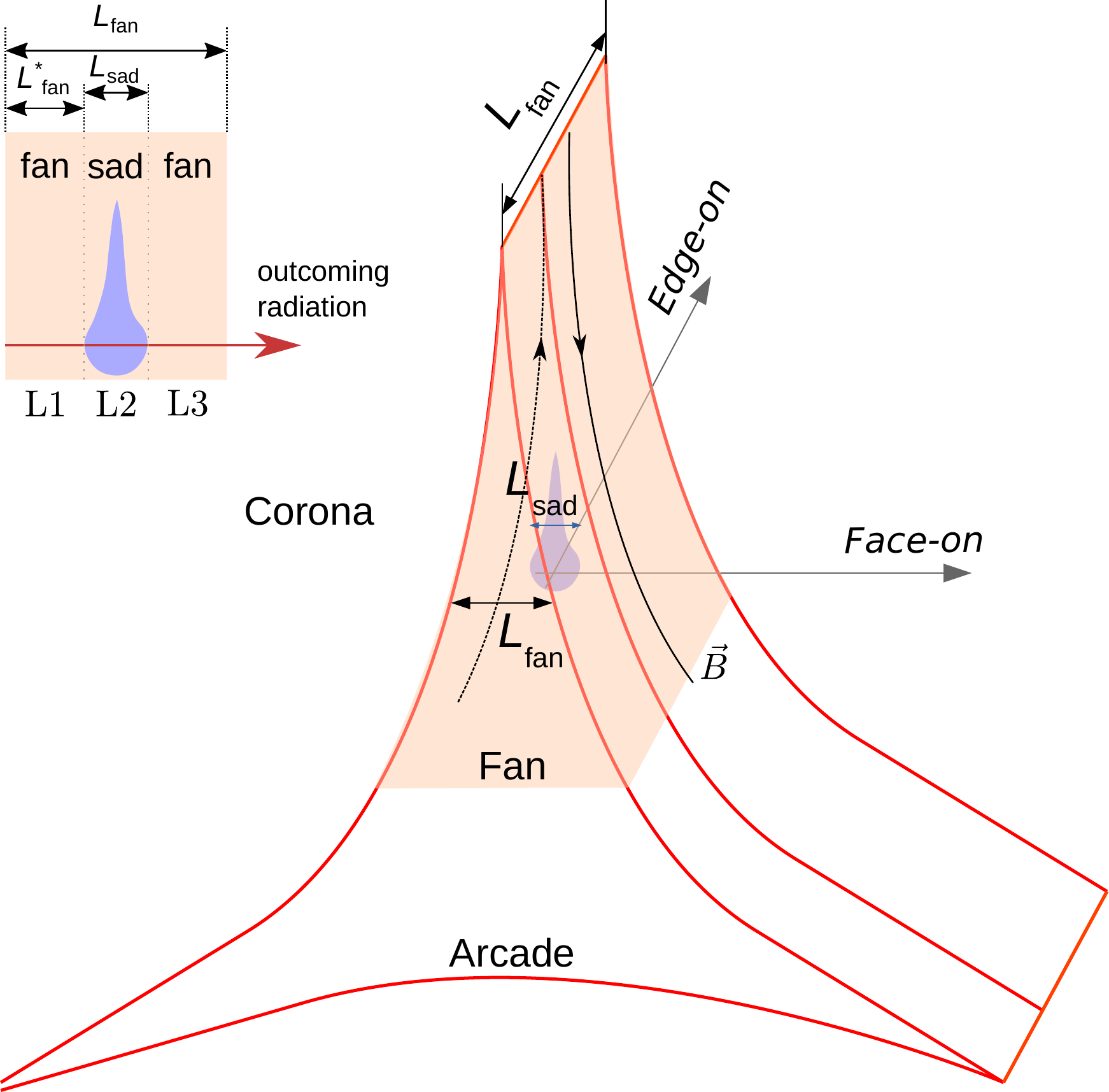}
\caption{Schematic scenario of a SAD (blue feature) embedded in the bright, dense plasma of the fan (represented by red lines). Below the fan lies the post-flare arcade, and the external coronal medium surrounds it. The fan additionally embeds the expected supra-arcade current-sheet. The fan is represented as viewed by an observer to the limb with two particular line-of-sights of interest represented, the edge-on and face-on views. Given that the fan is a thin, linear, long structure along the edge-on direction, when the fan is viewed edge-on the line-of-sight depth $L_\mathrm{{fan}}$ is greater than when is viewed face-on. \textbf{(Inset)} Sketches a three-layer scheme used to compute the outcoming radiation considering a SAD embedded in the fan for an observer in the arrow direction. We note that in general $L_\mathrm{fan}>L_\mathrm{sad}$.}
\label{fig:sketch}
\end{center}
\end{figure}

In EUV and soft X-ray observations coronal supra-arcade downflows (SADs) are seen as wiggly opaque structures leaving extended dark wakes during their descending motions toward the solar surface, to finally stop and gradually disappear. Sometimes their motions resemble tadpoles swimming upstream through the interstice of the fluid that opposes less resistance. SADs have usually been detected during early decay phases of long-term eruptive flares producing coronal mass ejections, and SADs' typical lifetime is of a few minutes. Today there is a general consensus that SADs are voided cavities of plasma, with densities lower than their surroundings and thus having an intrinsically lower brightness \citep{2003innesSoPh217.247}. SADs have been reported in several events \citep{1999mckenzieApJL519,2009mckenzieApJ697,2011savageApJ730,2011mackenzieApJL735,2011warrenApJ742,2014hannemanApJ786,2017chenA&A606} and with different instruments and observational techniques: direct images, with TRACE in EUV; XRT/Hinode and AIA/SDO in soft X-rays; and spectra with SUMER/SOHO in EUV. SADs are observed to descend from the upper part of the \textit{fan}, a region that lies above the arcade and is formed after the eruption, and which is also referred to in the literature as: \textit{post-eruption supra-arcade}, \textit{plasma sheet}; \textit{thermal halo}; \textit{supra-arcade fan}; hereafter referred as \textit{fan} (see sketch of Fig.~\ref{fig:sketch}). The fan is a dynamical region exhibiting turbulent motions with a relatively high plasma $\beta$ parameter \citep{2013mackenzieApJ766,2018freedApJ866}, is bright in EUV and soft X-ray wavelengths providing the necessary contrast for SADs detection, and is also hotter and denser than the external coronal medium by roughly an order of magnitude. Moreover, soft X-ray observations have provided evidence that the bright plasma of the fan surrounds a current sheet \citep{2013liuApJ767,2018warrenApJ854}, which, in line with the classical model of eruptive flares, it is expected to form behind the coronal mass ejection. It is assumed that the fan may only cover the lower part of the current sheet. On the other hand, SADs have been observed when the fan line-of-sight is over the limb, and preferably when viewed \textit{face-on}  perpendicular to the arcade axis (Fig.~\ref{fig:sketch}) during  the early decay phase after the soft X-ray peak flux of the erupting flare.

Despite that several models of SADs have been proposed, their origin and dynamics are not completely understood yet, therefore, what is their driving physical process  is still open to further research. Based on observations and/or theoretical considerations, a common point among all proposed models is the association of possible SADs' origins with magnetic reconnection processes occurring in the fan or in its upper coronal medium after the eruption (Fig.~\ref{fig:sketch}). Some models \citep{2009lintonEPS61,2009longcopeApJL690,2013scottApJ776,2014guoApJL796,2018longcopeApJ868} have pointed out to reconnection downflows descending through the fan as the driven mechanism: considering the standard model of erupting flares, behind the coronal mass ejection, magnetic field lines of opposite polarities are swept together forming the supra-arcade current sheet where reconnect. As a result, part of the reconnection outflows are propelled downward, via magnetic tension, to the underlying arcade, likely in the form of descending magnetic flux tubes or loops. \citet[][see also \citealt{2006lintonApJ642}]{2009lintonEPS61} carried out 3D MHD simulations considering an intermittent burst of patchy reconnection in a Y-type current sheet  supporting the idea that SADs are descending reconnected flux tubes. \cite{2012savageApJL747} interpreted SADs as subdense wakes left by shrinking reconnected flux tubes by the observational analysis of one event. And in turn, \citet[][see also \citealt{2014innesApJ796}]{2014guoApJL796} proposed that reconnected flux tubes in their retracting motions through the inhomogeneous fan generate interfaces of plasma where Rayleigh-Taylor instabilities are developed and subsequently generate the SADs' phenomenon. In these models, SADs are triggered in the current sheet neighbourhood, i.e. inside the fan or in its upper coronal region. A common issue with the reconnection downflow interpretation of SADs is the several reported discrepancy between the measured SADs speeds (often smaller than $200$~km s$^{-1}$) and the theoretical reconnection downflow speeds, expected to be the Alfv\'en speed ($\sim$1000~km s$^{-1}$) of the fan. Among possible explanations to overcome the speed mismatch, it was proposed that drag forces could act on the descending motion of reconnection downflows \citep{2018longcopeApJ868}, slowing them down. 

In contrast with the above reconnection downflow interpretation of SADs, \citet[][\citealt{2010schultzMNRAS407,2011maglioneAA527,2012cecereApJ759,2015cecereApJ807,2016zurbriggenApJ832}]{2009costaMNRAS400} assumed SADs to be generated by spontaneous, intermittent, bursty magnetic reconnection processes taking place somewhere in the fan or in its upper coronal region. One of this bursty reconnection event injects energy in a localised fast way, resembling a spark, and generating shock and rarefaction non-linear waves leading to the formation of a subdense cavity, whose structure supports the external pressure because is hotter than its surroundings, and which is then observed in EUV and soft X-rays as an opaque structure moving down along the bright fan. The spontaneous reconnection drivers are considered as local processes, which could be independent of the overall supra-arcade current sheet. A common characteristic of simulated SADs by \citet[][and their subsequent works]{2009costaMNRAS400} is that SADs are hotter than the fan. However, some observational studies \citep{2014hannemanApJ786,2017reevesApJ836,2020xueApJ898} have pointed out that SADs are hotter than the upper coronal medium but they do not seem to be hotter than the fan. This temperature discrepancy between SADs and the fan is due to simplifying assumptions on the modelling, which may be overcome by considering a more complex fan scenario. In addition, a questionable point related to the scenario proposed by  \citet{2016zurbriggenApJ832} is that simulated SADs exhibited a lack of elongated-tadpole-like shapes during their descends, having a more compact shape, and opposed to several observational reports. This issue is a consequence of considering explicitly anisotropic thermal conduction in the MHD simulations. 

Finally, given that SADs are not usually detected during eruptive flares, an important question still remaining to be answered by observations and models is whether SADs are an infrequent phenomenon, and thus rarely observed, or they are a common phenomenon usually not detected for some reason. SADs have been detected and mainly studied in EUV and soft X-ray frequencies, whereas in radio wavelengths only indirect evidences have been reported and, as far as we know, no attempt was made to directly detect them. In this work we compute the expected thermal bremsstrahlung emission produced by SADs in wavelengths from cm to submillimetre using the numerical MHD models presented by: \cite{2014guoApJL796}; \cite{2015cecereApJ807}; and \cite{2016zurbriggenApJ832}. Moreover, we synthesise maps by convolving telescope beams, of the Atacama Large Millimeter / Submillimeter Array (ALMA) in Chajnantor (Chile) and the Karl G. Jansky Very Large Array (VLA) in Socorro (USA), with the MHD simulations of SADs carried out by \cite{2016zurbriggenApJ832}. The goal of this work is to motivate and find out the best strategy to observe in cm\,--\,submillimetre wavelengths these illusive structures triggered during long-term solar erupting flares.

\section{Review of Observations} 
\label{sec:obs}                  
EUV and soft X-ray observations reported that the fan is hotter and denser than the background coronal plasma in approximately an order of magnitude, with fan temperatures in the $\sim$[10\,--\,20]~MK range and ion number densities of $\sim$[10$^9$\,--\,10$^{10}]$~cm$^{-3}$. In the literature, we find that SADs have been detected at heights of $\sim$[40\,--\,150]~Mm above the solar surface, with an average height of $\sim$80~Mm; they descend distances of $\sim$[10\,--\,20]~Mm; have downward speeds in the $\sim$[50\,--\,500]~km s$^{-1}$ range, but with a relatively  low average speed of $\sim$130~km s$^{-1}$; have sizes between $\sim$[1\,--\,10]~Mm; and their lifetime is around  $\sim$[3\,--\,5]~minutes, although there are few extreme exceptions observed during $\sim$12~minutes \citep{2014innesApJ796} and even for an hour \citep{2014hannemanApJ786}. Among all flare events reported in the literature with SADs detection or any signature of their presence, four of them were extensively studied: 21 April 2002 on NOAA active region (AR) 9906 \citep[e.g. see][]{2003innesSoPh217.267,2003innesSoPh217.247,2005verwichteA&A430}; 22 October 2011 on NOAA AR 11314 \citep{2013mackenzieApJ766,2014hannemanApJ786,2017reevesApJ836,2020xueApJ898,2021liApJ915}; 19 July 2012 on NOAA AR 11520 \citep{2013liuApJ767,2013liuMNRAS434,2014innesApJ796}; and 10 September 2017 on NOAA AR 12673 \citep{2018longcopeApJ868,2018warrenApJ854,2019hayesApJ875,2019caiMNRAS489,2020yuApJ900}. While the fan view for the first two events were face-on, the view for the third and fourth events were edge-on. On the other hand, the fan is a linear-like long column of plasma along the current sheet direction (edge-on view, see sketch of Fig.~\ref{fig:sketch}) with $\sim$[3\,--\,30]~Mm depths, but we note that it is considerable thinner in the perpendicular direction (face-on view). 

So far there have not been any direct SAD detection in radio or microwaves, at any frequency. On the other hand, some observational studies of flaring events \citep{2004asaiApJL605,2015chenSci350,2020yuApJ900}, occurring over or near the limb and preferably in edge-on views, revealed temporal and spatial correlations between EUV recurring downflow motions in the fan and subsequent detection of impulsive radio emission bursts on the top or leg of the lower arcade. In any case, the downflows had a counterpart in radio wavelengths. The above authors interpreted the radio impulsive emissions as a consequence of the SADs presence, assuming the reconnection downflow interpretation. 

\begin{table*}
    \begin{center}
    \begin{tabular}{|c|c|c|c|c|c|c|c|c|}
        \hline
        Model & $L_\mathrm{sad}$~[Mm] & $n_{\rm sad}$ & $\rho_{\rm sad}$~[g cm$^{-3}$] & $n_{\rm fan}$ & $\rho_{\rm fan}$~[g cm$^{-3}$] & $T_{\rm sad}$~[MK] & $T_{\rm fan}$~[MK] & $B_{\rm fan}$~[G] \\ 
        \hline
        1 & $2$    & $2.4$ & $5.1\times 10^{-15}$ & $5.8$ & $1.2\times 10^{-14}$ & 19.2 & 7  & 3   \\
        2 & $4$    & $8.9$ & $1.5\times 10^{-14}$ & $20$  & $3.3\times 10^{-14}$ & 22.4 & 10 & 5.9 \\
        3 & $12$   & $6.5$ & $1.1\times 10^{-14}$ & $20$  & $3.3\times 10^{-14}$ & 27   & 10 & 5.9 \\        
        4 & $4.7$  & $2.6$ & $4.3\times 10^{-15}$ & $3.3$ & $5.5\times 10^{-15}$  & 22   & 18 & 14  \\
        5 & $11.6$ & $2.6$ & $4.3\times 10^{-15}$ & $3.3$ & $5.5\times 10^{-15}$  & 22   & 18 & 14  \\
        \hline
    \end{tabular}
    \caption{Summary of the different numerical MHD models. $L_\mathrm{sad}$ is the SADs size, $n_\mathrm{sad}$ and $n_\mathrm{fan}$ the SADs (inner) and fan (outer) average ion number densities [cm$^{-3}\times 10^9$], respectively; $T_\mathrm{sad}$ and $T_\mathrm{fan}$ are the corresponding temperatures; and $B_\mathrm{fan}$ represents the mean magnetic field intensity of the fan. Data taken from the literature and  here referred as: model 1 from \citet{2016zurbriggenApJ832}; models 2\,--\,3 from \citet{2015cecereApJ807}; and models 4\,--\,5 from \citet{2014guoApJL796}.}
    \label{tbl:models}
    \end{center}
\end{table*}
%

\section{Modelling the expected emission} 
\label{sec:emission}                      

\subsection{SADs models description} 
\label{sec:models}                   

In order to calculate the thermal bremsstrahlung emissions of fans and SADs, we need to know their plasma temperatures and densities. Given that these data are not always available in the literature models, we circumscribe to the numerical MHD scenarios presented by: \citet{2014guoApJL796}; \citet{2015cecereApJ807}; and \citet{2016zurbriggenApJ832}. Table \ref{tbl:models} summarises the relevant mean characteristics of each model: SAD size $L_{\rm sad}$; SAD (inner) ion number density $n_{\rm sad}$ (also equivalent mass density $\rho_{\rm sad})$ and temperature $T_{\rm sad}$; and fan (outer) ion number density $n_{\rm fan}$, temperature $T_{\rm fan}$ and mean magnetic field intensity $B_\mathrm{fan}$. 

Following the work of \citet{2009costaMNRAS400}, and motivated by the turbulent description of the fan provided by \citet{2013mackenzieApJ766}, \citet{2015cecereApJ807} carried out 3D ideal MHD simulations to model a turbulent and dynamical fan as the medium where SADs descend. In order to generate a turbulent fan, a combination of tearing-mode and Kelvin-Helmholtz instabilities were used. It was argued that, to obtained SADs compatible with observations, the activation of spontaneous bursty reconnection processes are required, and that SADs support the external pressure because they are hotter than their surroundings. Moreover, for contrast requirements \citet{2015cecereApJ807} suggested that there must be a closed relation between SAD sizes and fan widths that should be satisfied for SADs to be observable. For this reason, SADs of two different characteristic sizes were generated: with $L_\mathrm{sad}=4$~Mm and $L_\mathrm{sad}=12~$Mm. 

\begin{figure}
\begin{center} 
\vspace{0.0 mm} 
\centerline{\includegraphics[width=0.495\textwidth]{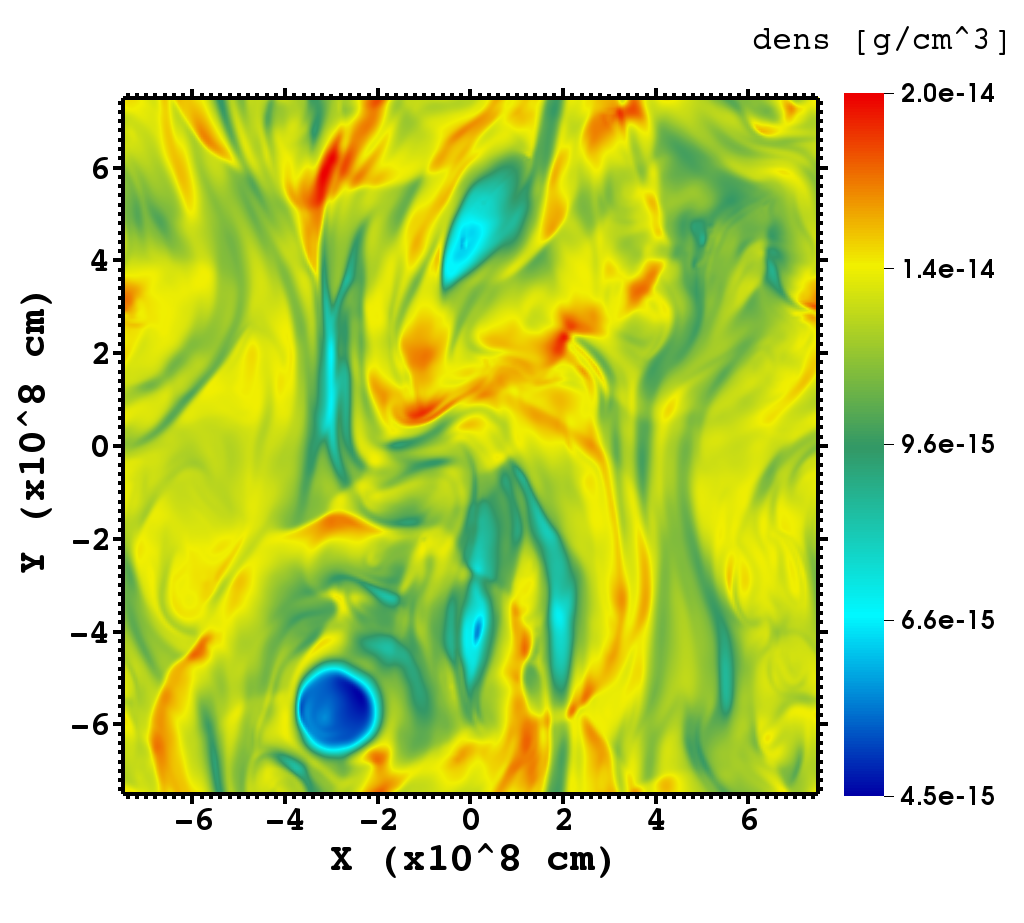}
            \includegraphics[width=0.495\textwidth]{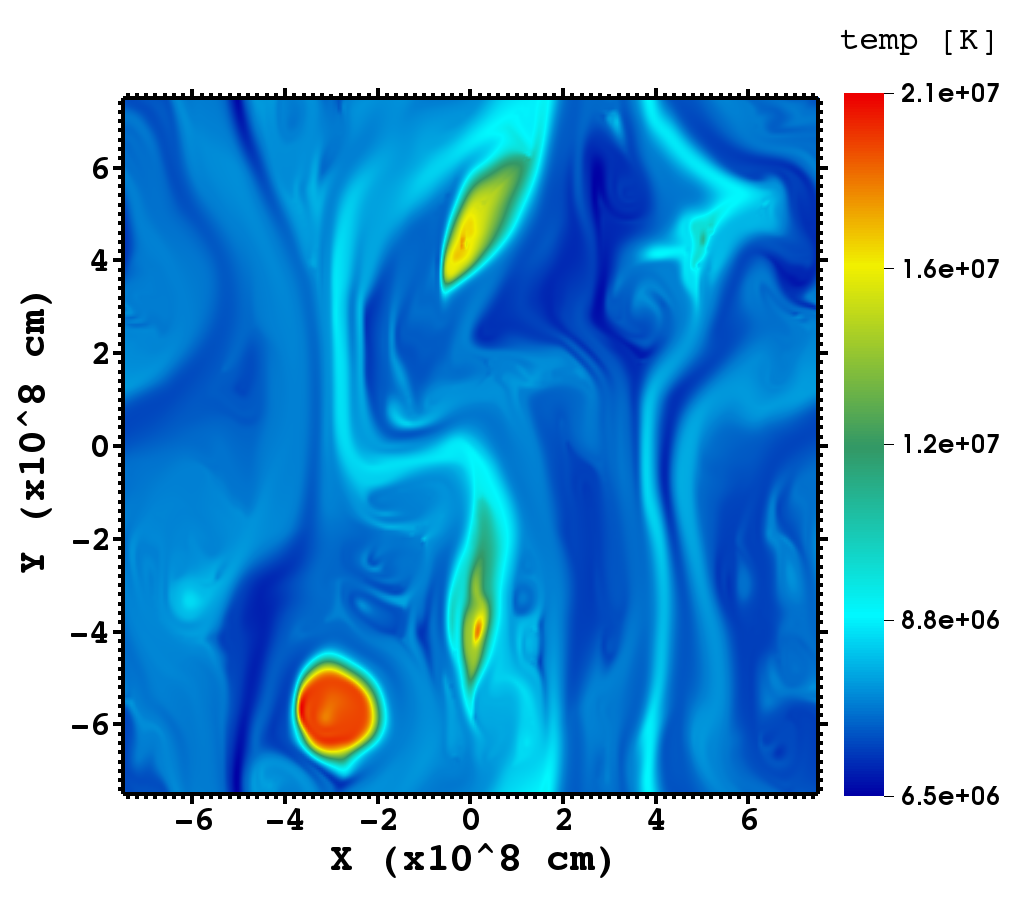}}
\caption{MHD numerical simulation of SADs descending through a turbulent fan explicitly considering thermal conduction \citep[presented in][]{2016zurbriggenApJ832}. Left: density values are shown. Right: corresponding temperature. This model represents a fan viewed face-on. Let us note $10^{8}~{\rm cm}\, \equiv 1~{\rm Mm}$.}
\label{fig:sads}
\end{center}
\end{figure} 

Considering the scenario presented by \citet{2015cecereApJ807}, the 2D MHD simulations of \citet{2016zurbriggenApJ832} took into account the particular case of small SADs ($L\approx 2~$Mm), where thermal conduction is expected to be more efficient and fade away SADs much faster than their typical lifetime. In this case a properly turbulent fan was generated using a stirring force source. Figure~\ref{fig:sads} shows images of density (left) and temperature (right) maps. The images display three SADs, two strongly faded away mainly due to the thermal diffusion, while the third one  at the bottom left corner with coordinates $(x,y)=(-3,-6)\times 10^8$~cm, and size $\approx$2~Mm $\equiv 2.8^{\prime\prime}$  is the most intense. This SAD has a mean density $\rho_{\rm sad} \approx 5.1\times 10^{-15}$~g cm$^{-3}$, while the fan is almost three times denser, $\rho_{\rm fan} \approx 1.2 \times 10^{-14}$~g cm$^{-3}$. On the other hand, the SAD inner temperature $T_{\rm sad} \approx 19.2$~MK is three times greater than the outer $T_{{\rm fan}} \approx 7$~MK. This model represents the case of a fan viewed face-on. 

\citet{2014guoApJL796}, performed 3D resistive MHD simulations to generate SADs by means of Rayleigh-Taylor instabilities associated to shrinking motions of magnetic flux tubes. In their scenario using constant magnetic resistivity, SADs have characteristic sizes within $[6^{\prime\prime}$\,--\,$15^{\prime\prime}] (\equiv [4.7$\,--\,$11.6]$~Mm, see their figure 2). The extreme values of this range were adopted in our calculations. The remaining values of density and temperature required for our analysis were extracted by visual inspection from their figures 2(b)\,--\,(c).

\subsection{Expected total flux} 
\label{sec:flux}

Table \ref{tbl:models} summarised the mean characteristics of the SADs and fans taken from the different numerical models discussed in Sect.~\ref{sec:models}. Hereafter we will refer to as model 1 to the smallest SAD (here considered) taken from \citet{2016zurbriggenApJ832}; models 2 and 3 to the small and big SADs simulated in \citet{2015cecereApJ807}; and models 4 and 5 to the extreme sizes ($6''$ and $15''$) reported in \citet{2014guoApJL796}. 

In order to compute the (free-free) thermal bremsstrahlung emission we use the expression 
\begin{equation}
F(\nu) = \frac{2k_{\rm B}T\nu^2}{c^2} \left(1 - e^{-\tau_\nu}\right) \Omega \ ,
\label{eq:flux}
\end{equation}
where $k_{\rm B}$ is the Boltzmann constant. $\tau_\nu$ is the optical depth at frequency $\nu$, which is approximated by \citep{1985dulkARAA23} 
\begin{eqnarray}
\tau_\nu &=& \kappa_\nu L \ , \nonumber \\
\kappa_\nu &=& 9.78\times 10^{-3} \frac{n_{\rm e}}{\nu^2 T^{3/2}} \left (\sum_{\rm i} Z_{\rm i}^2n_{\rm i}\right) \ G(T,\nu) \ \label{eq:kappa},
\end{eqnarray}
where $\kappa_\nu$ is the thermal bremsstrahlung opacity and $L$ is the plasma line-of-sight depth considered and $G(T,\nu)$ is the Gaunt factor that we obtain numerically from \cite{vanHoofetal:2014}. Also, $n_{\rm e}$ and $n_{\rm i}$ are the number densities of free electrons and ions of specie i, respectively, and $Z_{\rm i}$ is the atomic number of specie i. We consider a fully ionised, ideal plasma with a solar abundance, i.e. $70.7\,\%\, {\rm H} + 27.4\,\%\, {\rm He} + 1.9\,\%$ heavier elements \citep{2000prialnikBook}, implying that $\sum_{\rm i} Z_{\rm i}^2 n_{\rm i} = 3.703 n$, with $n$ the average total ion number density, and from which we derive the electron density $n_{\rm e}=1.445n$. $\Omega$ is the solid angle considered. 

\begin{figure*}
\begin{center}
\includegraphics[width=\textwidth]{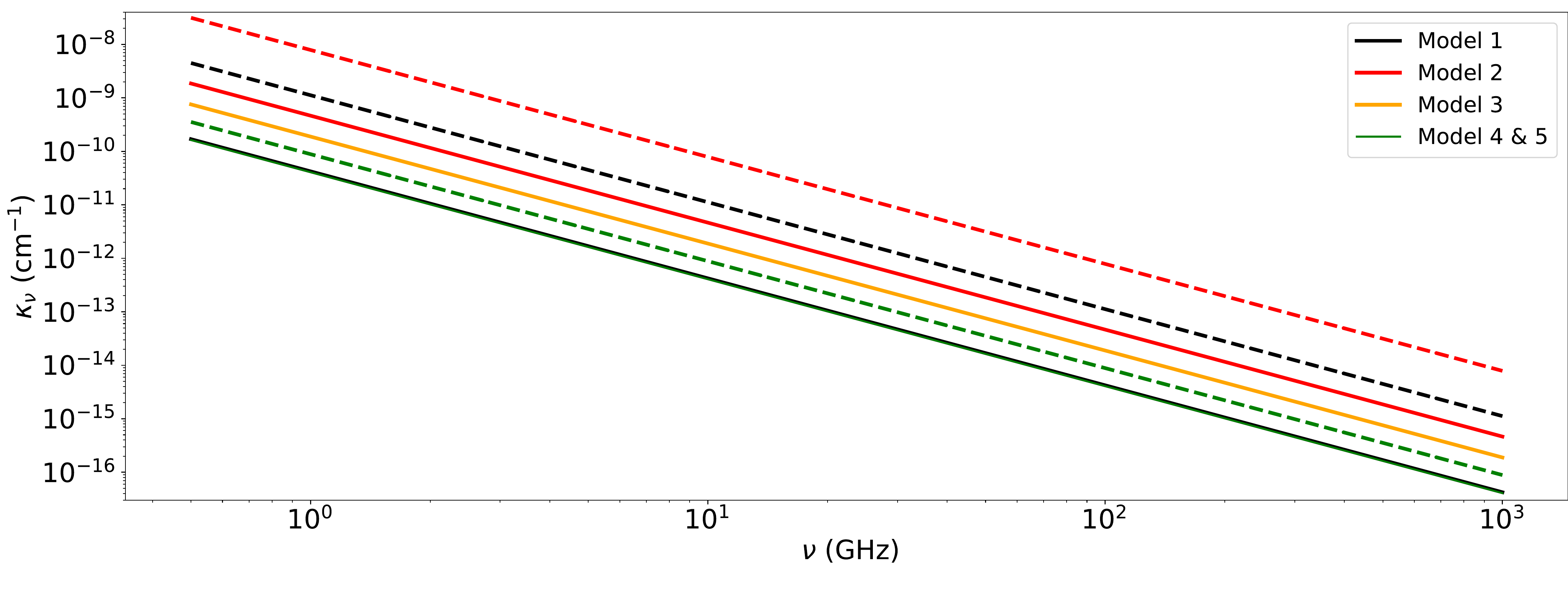}
\caption{Mean thermal bremsstrahlung opacities $\kappa_\nu$ for the different models (identified by colours) in the frequency $[0.5$\,--$\,1000]$~GHz range. Continuous lines correspond to inner (SAD) opacities, while dashed lines to outer (fan) opacities. Fan opacities for models 2 and 3 are equal; the same happens for  models 4 and 5. We note that SAD opacities for models 1, 4, and 5 are similar (lowest superimposed continuous lines).}
\label{fig:kappa}
\end{center}
\end{figure*}

From Eq.~(\ref{eq:kappa}) we compute the thermal bremsstrahlung opacities $\kappa_\nu$ as a function of the observing frequency $\nu$ for the SADs and fans of models 1\,--\,5. Figure~\ref{fig:kappa} shows the obtained results. Colour-continuous lines represent the SAD opacities, while colour-dashed lines the corresponding fan opacities which in general are larger in values. For fans and SADs the regime should be optically thin above $10$~GHz. However, in some cases that will be analysed later, the transition from optically thick to thin regimes may happen at frequencies below $1$~GHz.

We can now estimate the integrated flux density emitted by the whole SAD using  mean values from Table~\ref{tbl:models} and using Eq.~(\ref{eq:flux}). The intrinsic turbulent dynamics of the fan  implies that random fluctuations cancel out when integrated over the line of sight, and therefore the use of mean values is appropriate for computing  outgoing radiation observed on Earth. In the calculation we assume that all sources have a cylindrical shape with a diameter equal to the SAD size, and the solid angle $\Omega = \pi L_{\rm sad}^2/4 {\rm AU}^2$, with ${\rm AU}$ the mean Sun\,--\,Earth distance. Because of we are considering an observer viewing the fan over the limb, the chromospheric contribution is not taken into account. As sketched in Fig.~\ref{fig:sketch}, we assume an observer in the outcoming radiation direction and that the SAD is embedded in the fan, thus we take into account the fan emission in front of and behind the SAD. Given that we are interested in the contrast, we want to compare the emissions coming from a direction including the SAD with another one just considering the surrounding fan. When the considered direction include the SAD, a three-layers calculation is done (see the inset in Fig.~\ref{fig:sketch}), otherwise just one layer is taken into account. For all cases the layers are assumed isothermal and homogeneous. In the three-layers calculation of the flux density, the inner (L1) and outer (L3) layers are formed by the fan plasma, and the middle layer corresponds to the SAD (L2). Here we consider that the SAD layer is in the geometrical centre of the fan, and we define as $L^*_{\rm fan} = (L_{\rm fan} - L_{\rm sad})/2$ the depth of the inner and outer layers. In this case the final total flux density is then

\begin{equation}
    F_{\rm sad}(\nu) = F_{\rm L1}(\nu) + F_{\rm L2}(\nu) + F_{\rm L3}(\nu) \ .
\label{eq:in}
\end{equation}
The emerging flux density of L1, after traversing the SAD and the outer fan layers, is
\begin{equation}
F_\mathrm{L1}(\nu) = \frac{2k_{\rm B}T_\mathrm{fan}\nu^2}{c^2} 
\left(1 - e^{-\tau_\mathrm{L1}(\nu)}\right) e^{-\tau_\mathrm{L2}(\nu)} e^{-\tau_\mathrm{L3}(\nu)}\Omega \ ,
\end{equation}
where $\tau_\mathrm{L1}$, $\tau_\mathrm{L2}$, and $\tau_\mathrm{L3}$ are the opacities of L2 and L3, respectively. In a similar way, we solve the radiation transfer equation for the SAD (L2) layer
\begin{equation}
F_\mathrm{L2}(\nu) = \frac{2k_{\rm B}T_\mathrm{sad}\nu^2}{c^2} \left(1 - e^{-\tau_\mathrm{L2}(\nu)}\right)  e^{-\tau_\mathrm{L3}(\nu)}\Omega \ .
\end{equation}
And for the outer layer we have
\begin{equation}
F_\mathrm{L3}(\nu) = \frac{2k_{\rm B}T_\mathrm{fan}\nu^2}{c^2} \left(1 - e^{-\tau_\mathrm{L3}(\nu)}\right) \Omega \ .
\end{equation}
On the other hand, the emerging radiation when observing off the SAD direction has only the fan contribution (a one-layer calculation)
\begin{equation}
F_\mathrm{fan}(\nu) = \frac{2k_{\rm B}T_\mathrm{fan}\nu^2}{c^2} \left(1 - e^{-\tau_\mathrm{fan}(\nu)}\right) \Omega \ ,
\label{eq:out}
\end{equation}
where $\tau_\mathrm{fan}(\nu)$ is the opacity of the whole fan--line-of-sight depth $L_\mathrm{fan}$. 
\begin{figure} 
\begin{center}
\vspace{-3.5 mm}
\centerline{\includegraphics[width=0.45\textwidth]{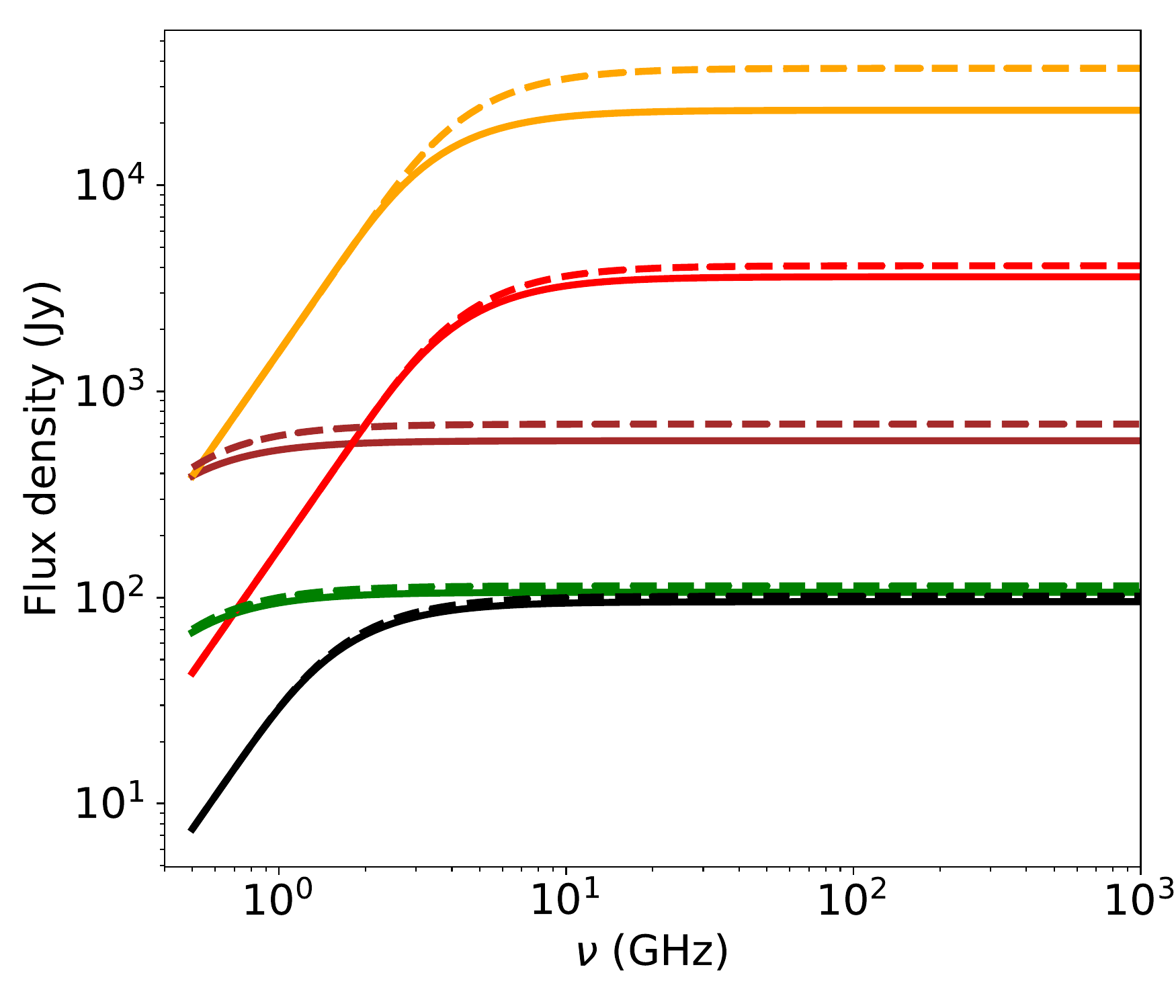}
            \includegraphics[width=0.45\textwidth]{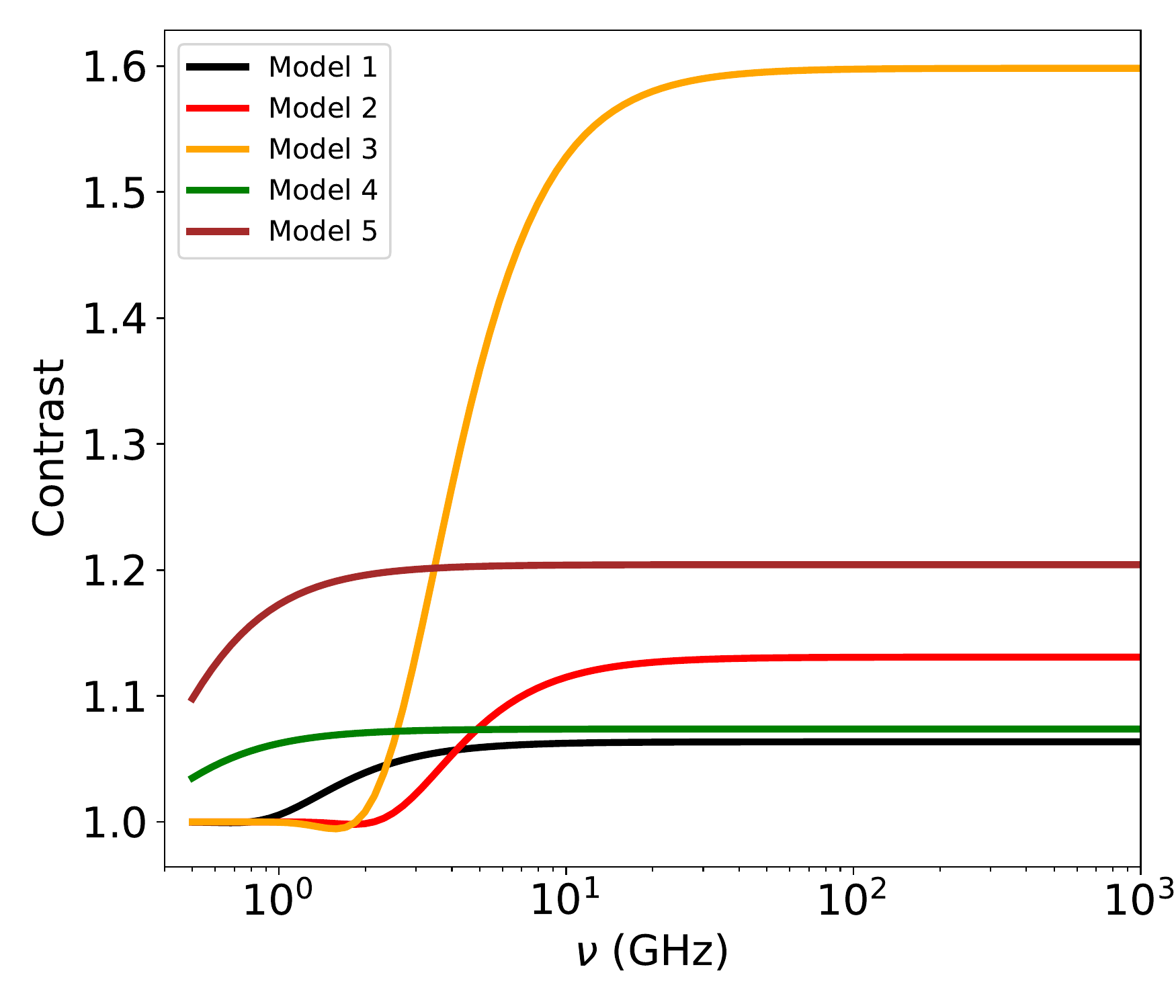}}
\vspace{-1.0 mm}
\centerline{\includegraphics[width=0.45\textwidth]{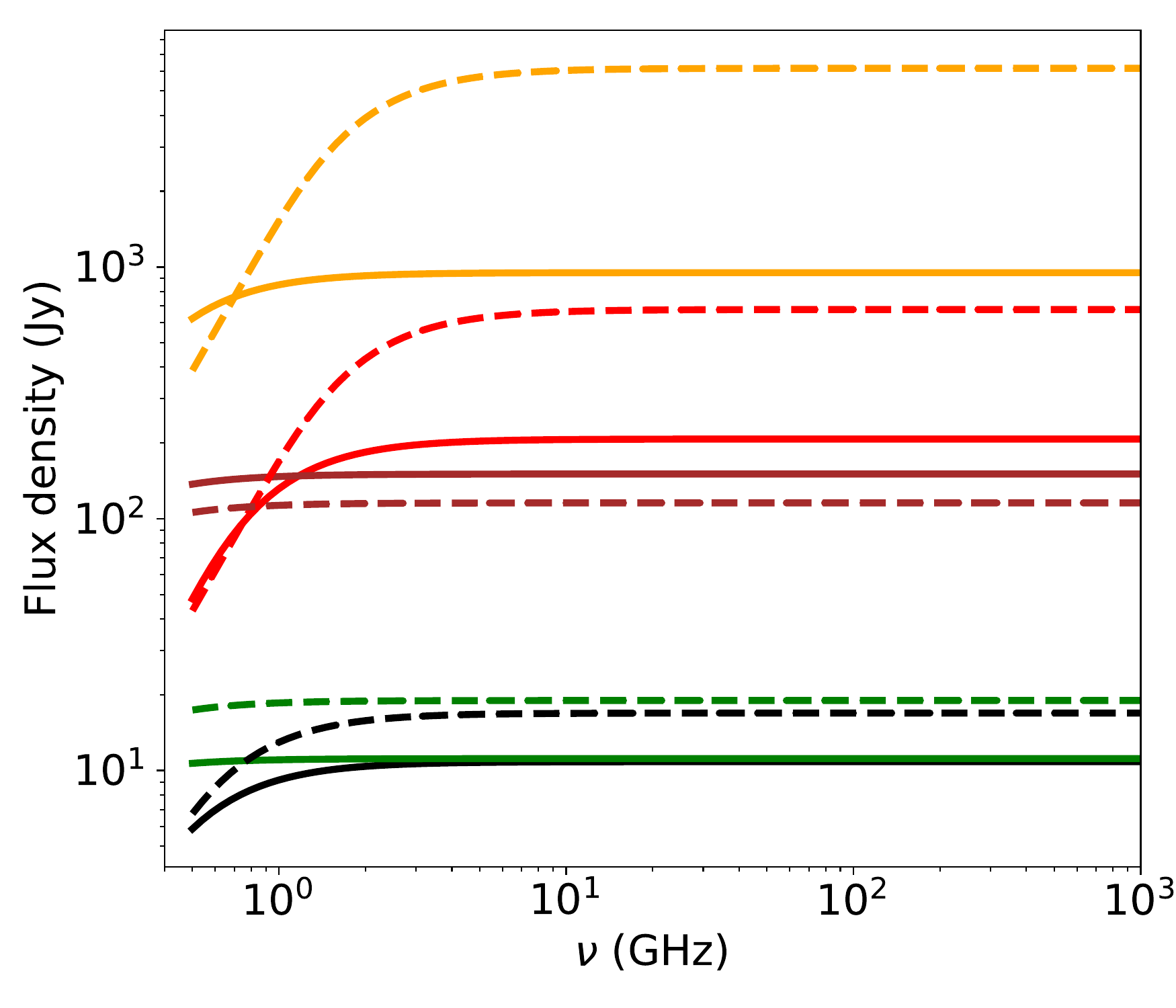}
            \includegraphics[width=0.45\textwidth]{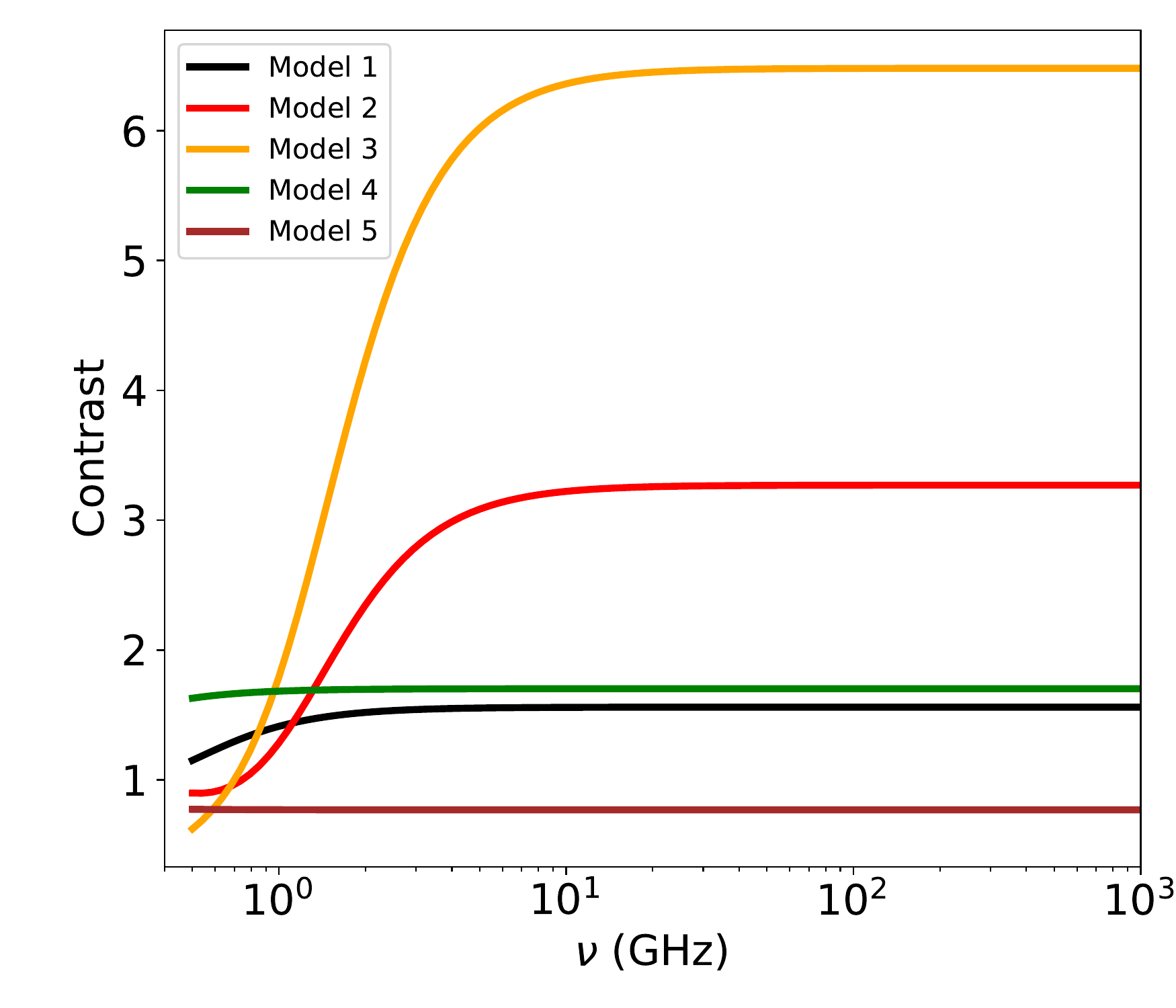}}
\caption{Top row: results for the fan viewed edge-on. Bottom row: results for the fan viewed face-on. Left: Expected flux density spectra. Continuous lines represent the flux density incoming from the SAD, while dashed lines the fan flux density. Right: contrast, flux density quotient between the fan and SAD.}
\label{fig:Fx}
\end{center}
\end{figure}

Figure~\ref{fig:Fx} shows the resulting integrated flux densities (top and bottom left) and the contrasts $Q = F_{\rm fan}(\nu)/F_{\rm sad}(\nu)$, dividing Eqs.~(\ref{eq:out}) by (\ref{eq:in}) (top and bottom right). Results are displayed for the edge-on and face-on views of the fan, with line-of-sight depths $L_\mathrm{fan}=30$~Mm and $L_\mathrm{fan}=5$~Mm, respectively. We note that the flux density is almost constant for frequencies $\nu\gtrsim 5$~GHz, where the spectrum is optically thin. The integrated SAD flux densities $F_\mathrm{sad}(\nu)$ are in the range $\sim$[10$^2$\,--\,$10^4]$~Jy, and are weaker than the fan flux densities $F_\mathrm{fan}(\nu)$. This is also evident in the right panels where we see contrasts $Q>1$. Model 3 exhibits the highest contrasts, 1.6 on the edge-on view and 6 on the face-on view, and model 1 the lowest, with 1.05 and 1.7, respectively, whereas the remaining contrasts lie between these extremes. When the spectra become optically thick ($\nu\lesssim 5$~GHz), the contrasts tend to unity, in particular in the edge-on view, meaning that the SAD layer contribution to the total flux density is negligible. Interestingly, for model 3 when observing the fan face-on, we see $Q<1$, implying that the SAD is brighter than the fan.  

\subsection{Gyroresonance}

In the previous calculations we did not take into account the gyroresonance radiation, an emission mechanism that depends on the magnetic field intensity, besides temperature and density. Gyroresonance opacity at frequency $\nu$ depends on the harmonic number  $s = \nu/\nu_B$, where $\nu_B = 2.8 \times 10^6 B$, with $B$ the magnetic field intensity in G. The opacity at a given harmonic number $\tau_s$ \citep[e.g. ][]{Casinietal:2017} is proportional to
$\tau_s \propto \frac{s^2}{s!}$. Because of this dependence only the very first harmonics produce detectable emission. Indeed, usually only the first three harmonics are taken into account \citep[e.g.][]{White:2004,Selhorstetal:2008}. In the present case, where magnetic field intensities of the MHD simulations are $B\le 14$~G,  $\nu_B \le 0.039$~GHz, therefore for the lowest frequency $\nu=0.5$~GHz we have $s \ge 13$, and $\tau_s \propto s^2/s! \lesssim 10^{-8}$. Since for higher frequencies the harmonic number $s$ will be even larger, we can neglect the gyroresonance contribution to the emerging flux density in the frequency range $[0.5$\,--$\,1000]$~GHz used in our calculations. 
\begin{figure*}
\begin{center} 
\vspace{-3.5 mm}
\centerline{\includegraphics[width=0.48\textwidth]{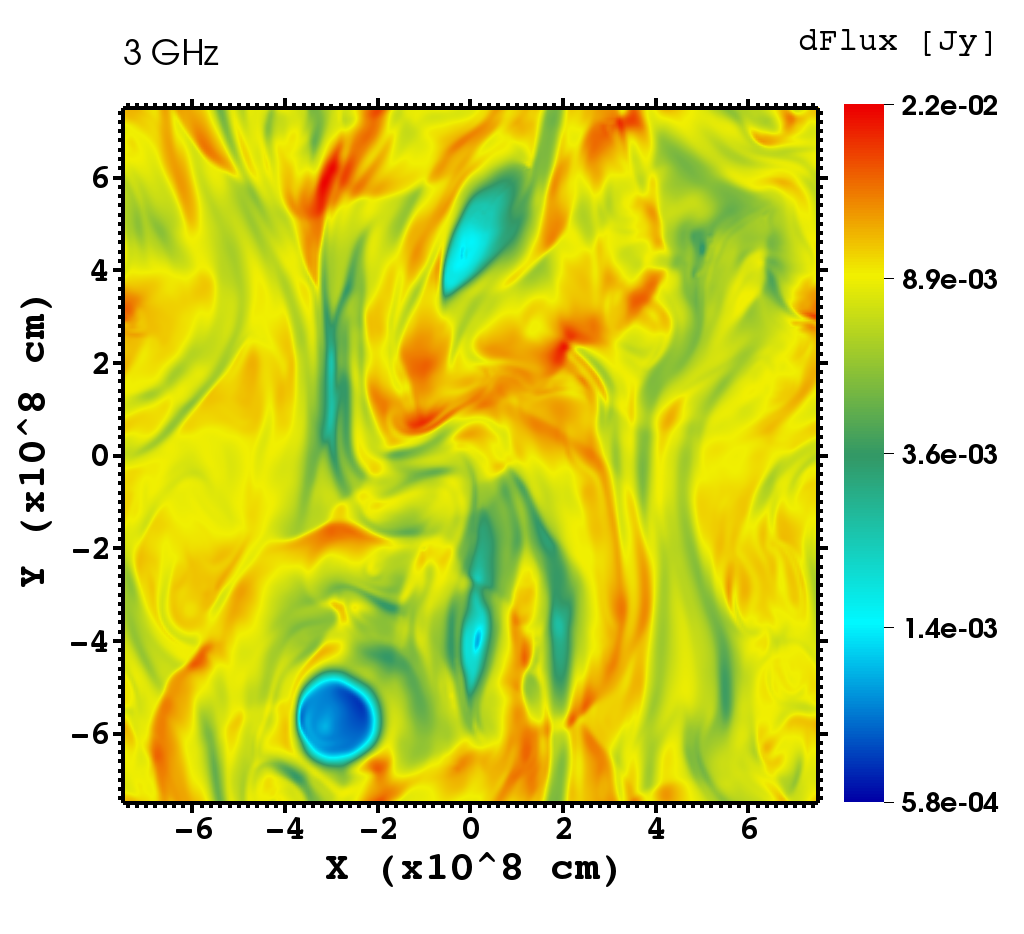}
            \includegraphics[width=0.48\textwidth]{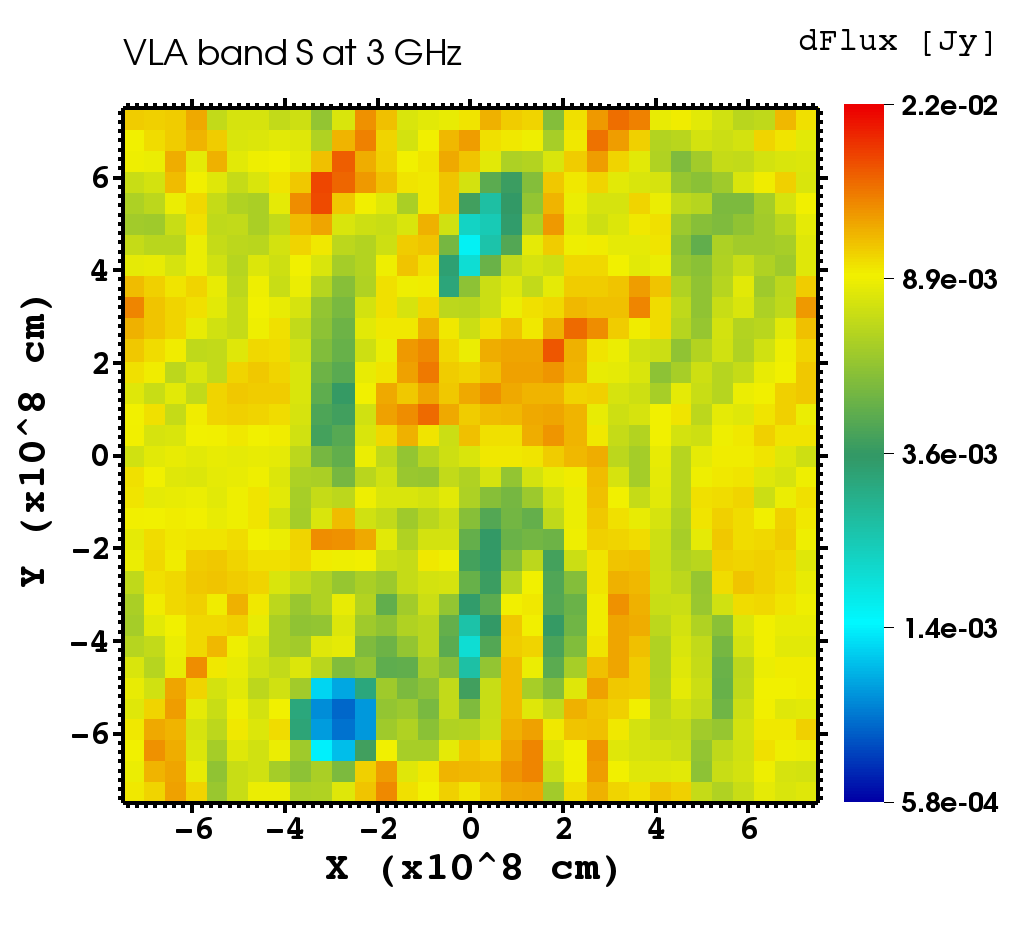}}
\vspace{-3.5 mm}
\centerline{\includegraphics[width=0.48\textwidth]{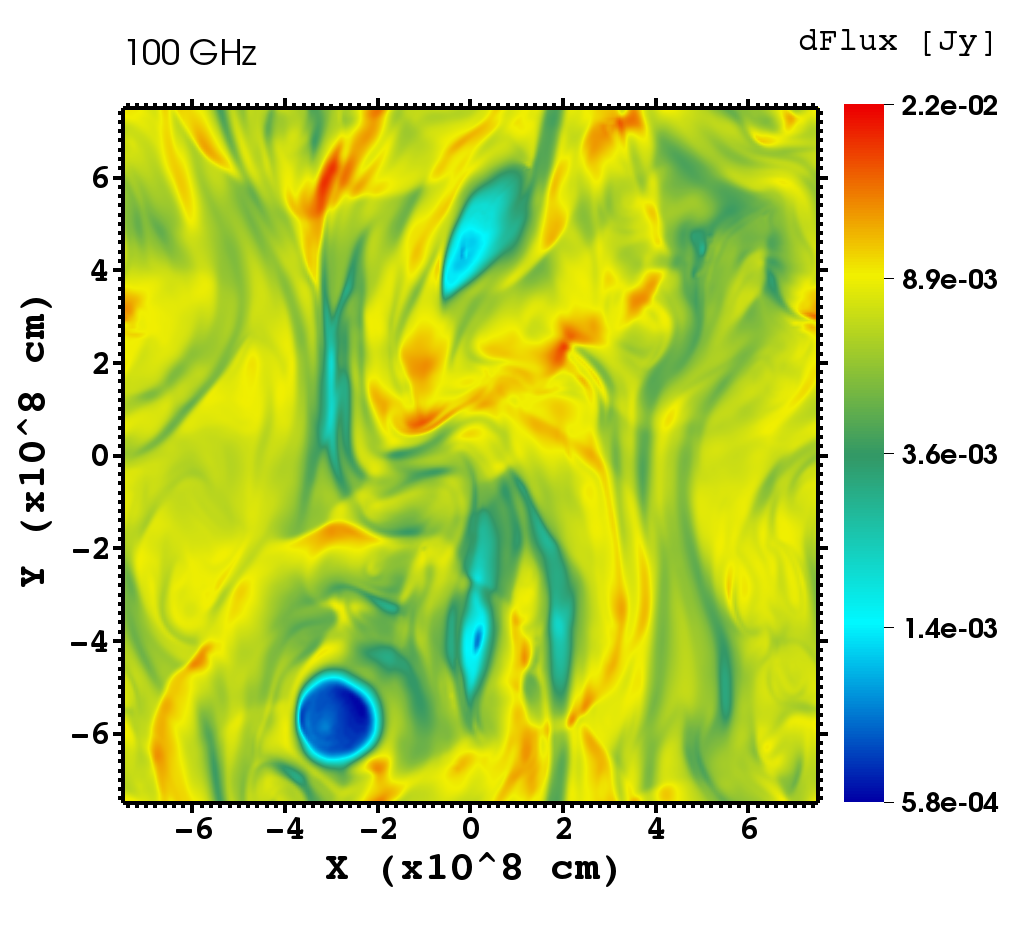}
            \includegraphics[width=0.48\textwidth]{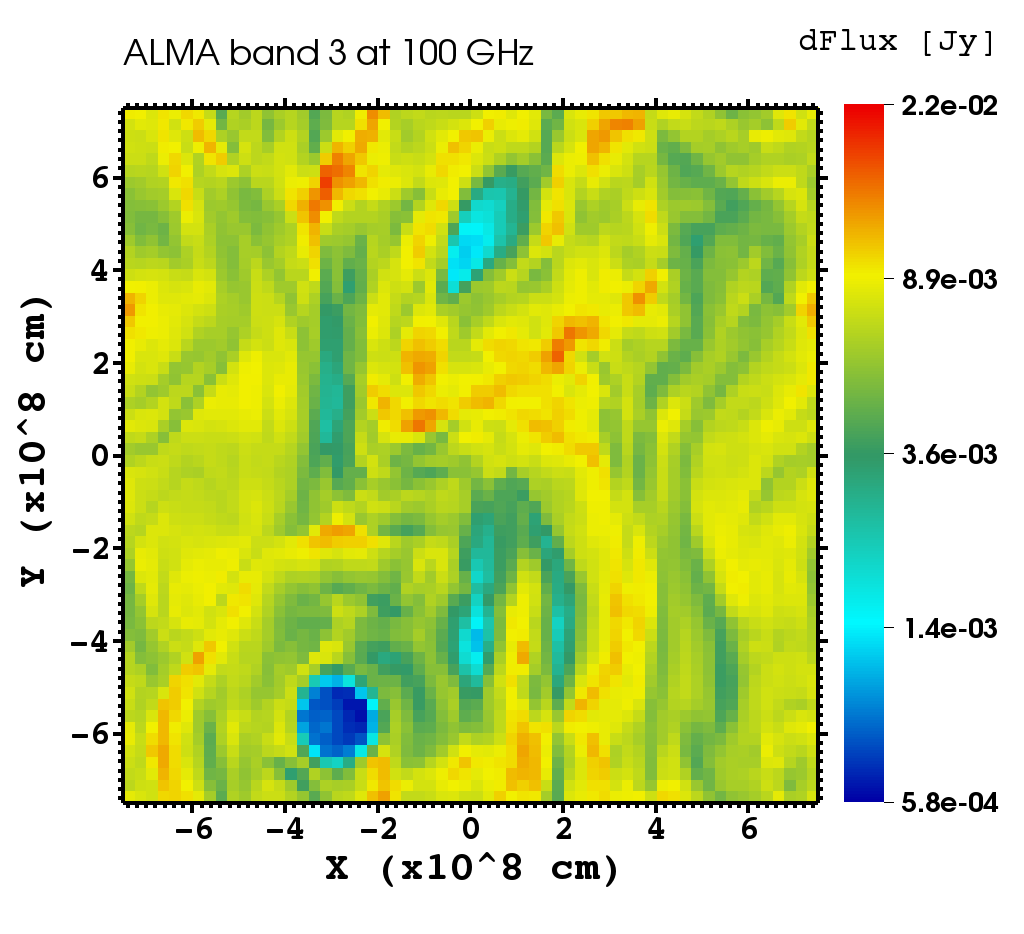}}
\caption{Top row: expected flux density at 3 GHz (left), and the flux density convolved with the VLA beam of ${\rm HPBW}=0.65^{\prime\prime}$ (right). Bottom row: expected flux density at 100 GHz (left), and the flux density convolved with the ALMA-band 3 beam of ${\rm HPBW}=0.3^{\prime\prime}$ (right). Let us note $10^{8}~{\rm cm}\, \equiv 1~{\rm Mm}$.} 
\label{fig:maps}
\end{center}
\end{figure*}
%

\subsection{2D synthesised radio maps} 
\label{sec:2Dmaps}                     

Here we use model 1 to synthesise images that we expect to observe with present radiotelescopes. Since model 1 is 2D, we do not include the fan contribution in front of the SAD or behind it as done above in Section \ref{sec:flux}, instead every grid cell is thought as a homogeneous source with a line-of-sight depth $L=5$~Mm. This value for the line-of-sight depth comes from the fact that model 1 accounts for a fan viewed face-on. The solid angle $\Omega=\delta x^2/{\rm AU}^2$ is considered, which is defined by the grid resolution ($\delta x = \delta y = 5\times 10^6$~cm) of the simulation. Using Eq.~(\ref{eq:flux}) we obtain the flux density emitted by every grid cell, and then to simulate observations, the resulting flux map is convolved with representations of instrument beams. In this case, due to the small size of the SAD, an instrument with high spatial resolution is needed to separate the SAD emission from the background fan. Nowadays there are only two instruments with arcsec spatial resolution in radio wavelengths and capable of observing the Sun: the VLA and ALMA. Both are interferometers with synthesised beams of the order of $1$~arcsec and sensitivities below 1 mJy.

We compute the emission for the VLA microwave band S ($3$~GHz), and also for the ALMA bands: 3 ($100$~GHz), 6 ($230$~GHz), and 9 ($720$~GHz). Since the spectrum is mostly optically thin, and aiming to display results for a low and a high frequency in the thermal-bremsstrahlung-opacity domain considered in Fig.~\ref{fig:kappa}, we show in Fig.~\ref{fig:maps} synthesised maps for the VLA band S at $3$~GHz (top row), and the ALMA band 3 at $100\,$GHz (bottom row). Left panels represent the flux densities with the simulation-grid spatial resolution, while the right panels show the resulting flux densities convolved with the two instrument beams, at the top using the VLA band S with a half power beam width ${\rm HPBW}=0.65^{\prime\prime}$, and at the bottom using the ALMA band 3 with ${\rm HPBW}=0.3^{\prime\prime}$. We note that the fan emission is always brighter than the SAD, with a contrast $Q \approx 6$. On the other hand, the convolved maps depend more on the spatial resolution (instrument beam) than on the flux density. For small SADs such as that of model 1, to use instruments with ${\rm HPBW} > 1^{\prime\prime}$ will result in SADs spatially unresolved from the background fan emission. In terms of sensitivity, the expected flux detected by the instruments is on the order or above of $1$~mJy/pixel.

In the 2D scenario, the order of magnitude for the emission contrast $Q$, between the fan and SADs, can be estimated straightforwardly in the optically thin regime. As it was pointed out above, the contrast depends on the model considered. E.g. for models 1\,--\,3, the fan temperatures are roughly half of the SADs ones, whereas the fan densities are roughly twice of the SADs ones. Due to we are considering the optically thin regime: the opacity $\kappa_\nu \propto n^2 \nu^{-2} T^{-3/2}$ (Eq.~(\ref{eq:kappa})); furthermore, the emission in the Rayleigh-Jeans approximation can be written as $F_\nu \propto T\nu^2\kappa_\nu L \Omega \equiv n^2 T^{-1/2} L \Omega$; and thus, for these models we obtain an order-of-magnitude contrast
\begin{equation}
    Q = \frac{n_{\rm fan}^2 \ T_{\rm fan}^{-1/2} \ L \ \Omega} {n_{\rm sad}^2 \ T_{\rm sad}^{-1/2} \ L \ \Omega} 
    = \left (\frac{n_{\rm fan}}{n_{\rm sad} }\right)^2 \left( \frac{T_{\rm sad}}{T_{\rm fan}}\right)^{1/2} \simeq 2^2 \cdot 2^{1/2} = 5.6 \ .
    \label{eq:contrast}
\end{equation}
This number is consistent with the contrast displayed by the flux densities in Fig.~\ref{fig:maps}. 
In addition, and as a support to the 2D-scenario results, the contrast of Eq.~\ref{eq:contrast} is of the same order as those reported by \cite{2014guoApJL796} in their figures~2(d)\,--\,(e), where 2D slices of the expected emissions were shown corresponding to face-on views of the SDO/AIA Fe\,XXI and Fe\,XXIV channels, respectively. Finally, if we repeat the simple contrast calculation of Eq.~\ref{eq:contrast}, but instead using the 2D-slice density and temperature found by \citet{2015cecereApJ807} in their figure~5, we again obtain a similar value. However, the contrast decrease considerable when the plasma line-of-sight depth is taken into account, as illustrated by Fig.~\ref{fig:Fx}.

\subsection{Observational strategies} 
\label{sec:strategies}                

It is evident from the flux density spectra displayed in Fig.~\ref{fig:Fx} that observations with edge-on views produce lower contrasts than with face-on views, making more difficult the SADs detection, a fact that worsens in the optically thick regime. Since the emission from microwaves to submillimetre wavelengths is mostly optically thin, the best strategy to observe SADs is when the fan line-of-sight is over the dark limb, and furthermore, when the fan is viewed face-on. On the other hand, according to the models analysed in this work, the spatially integrated flux emitted by SADs can be as low as $\approx$1~Jy and as high as $\approx$10$^3$~Jy. This means that it is not required a highly sensitive telescope, although the imaging capability is needed. The best instrument is an interferometre, e.g.: VLA has receivers in the low frequency domain $[1$\,--\,$18]$~GHz; and ALMA has receivers in the high frequency domain $[45$\,--\,$900]$~GHz. Both instruments have adequate sensitivity and spatial resolution to observe even the small SAD ($\approx$3~arcsec) of model 1, shown in Figs.~\ref{fig:sads}, \ref{fig:Fx} and \ref{fig:maps}. Therefore, the best strategy would be to observe at different frequencies simultaneously with VLA and ALMA to cover the possible maximum frequency range. Moreover, instruments below the $1$~GHz domain, like Low Frequency Array \citep[LOFAR,][]{vanHaarlemetal:2013} and Nan\c{c}ay Radioheliograph \citep[NRH,][]{KerdraonDelouis:1997}, can supply information in the optically thick spectral range, allowing a better determination of the density and temperature of the emitting sources. The Extended Owens Valley Array \citep[EOVSA,][]{2018garyApJ863}, with its spectral imager in $[1$\,--\,$18]$~GHz range could be used for the biggest SADs, since its beam sizes are of the order of $\approx$54$/\nu [\mathrm{GHz}]$~arcsec.

\section{Final remarks}  
\label{sec:end}          
So far, SADs have been elusive structures, likely given rise to by magnetic reconnection processes during solar flares. They have been detected only in EUV and soft X-ray emission, implying that currently known SADs characteristics, such as their temperatures and sizes, have been inferred using this wavelength range.  The same thermal bremsstrahlung phenomenon that produces emission in EUV and soft X-ray should also produce emission in radio wavelengths. However, no direct observation of SADs in radio have been reported in the literature. Instead, only signatures of magnetic reconnection processes assumed to be responsible for SADs triggering, and also their spatial/temporal correlations with SADs signatures, were reported. Thus, the use of radio observations for SADs detection may help to improve the estimations of SADs characteristics.

In order to contribute to the understanding of SADs' nature, in this paper we produced spatially integrated spectra, and synthesised images at selected radio frequencies. The spectra were obtained utilising numerical MHD models of SADs based on the bursty localised reconnection interpretation \citep[][and subsequent works]{2009costaMNRAS400}, and the reconnection downflow interpretation \citep{2014guoApJL796}. In the frequency range analysed, $[0.5$\,--\,$1000]$~GHz, the spectra of SADs and fans show that the thermal bremsstrahlung emission above $10$~GHz is mostly optically thin, and therefore proportional to: the temperature, the density, and the line-of-sight depth. Also the gyroresonance
contribution to the flux density is negligible. The fan emission and SADs sizes have a strong influence in their measured contrast $Q$. The contrast is a parameter that describes the \textit{detectability} of a SAD, meaning that it is indistinguishable from the ambient-fan emission when it is close to unity. The fact that the contrast is not sufficiently high may be a reason why SADs are barely detected during eruptive events, in particular when the fan is viewed edge-on (as shown in Fig.~\ref{fig:Fx}). 

In spite of our results are model dependent, a strategic plan to increase chances of SADs detection in radio wavelengths is, first, to search for post-flare magnetic arcades over the limb, where a face-on view of the fan will be mostly desired, and second, to simultaneously observe at cm and mm wavelengths. Imaging SADs with arcsec resolution simultaneously at different radio wavelengths would be an excellent diagnostic to constraint models, being the best instruments to achieve this goal the VLA and ALMA interferometers. Moreover, a combination of radio, EUV, and soft X-ray observations will enrich the description and deepen our understanding of the physical processes. 

It is known that magnetic reconnection is in the origin of a great variety of phenomena: flares, CMEs, etc. Some of these phenomena are weak and therefore difficult to detect. SADs are a phenomenon belonging to this category. In order to fully understand the reconnection process and its consequences, we need to complete the SADs' picture. Radio observations can be a key to have a more detailed picture of them.

\section*{Acknowledgments}
EZ is grateful to the FAPESP to have financed this research by the grant 2018/25177-4. GGC thanks CNPq for support with a Productivity Research Fellowship. The research leading to these results has received funding from CAPES grant 88881.310386/2018-01, FAPESP grant 2013/24155-3. The authors are also grateful with Mackenzie Research Funding Mackpesquisa for the received support. GGC is Correspondent Researcher of the Consejo Nacional de Investigaciones Cient\'ificas y T\'ecnicas (CONICET) for the Instituto de Astronom\'ia y F\'isica del Espacio (IAFE), Argentina. MC and A. Costa are members of the Carrera del Investigador Cient\'ifico (CONICET). MC acknowledges support from ANPCyT under the grant PICT No. 2016-2480. MC also acknowledges support from the SECYT-UNC grant No. 33620180101147CB. We also thank the {\sc VisIt} team for developing the graphical tool used in this work \citep{2012harrison2012python}. Part of the MHD numerical simulations here presented were performed with the {\sc Flash} code: ``The software used in this work was developed in part by the DOE NNSA ASC- and DOE Office of Science ASCR-supported Flash Center for Computational Science at the University of Chicago''.

\bibliographystyle{frontiersinSCNS_ENG_HUMS}  
\bibliography{references}

\begin{thebibliography}{47}
\providecommand{\natexlab}[1]{#1}
\expandafter\ifx\csname urlstyle\endcsname\relax
  \providecommand{\doi}[1]{doi:\discretionary{}{}{}#1}\else
  \providecommand{\doi}{doi:\discretionary{}{}{}\begingroup
  \urlstyle{rm}\Url}\fi
\providecommand{\selectlanguage}[1]{\relax}
\providecommand{\bibAnnoteFile}[1]{%
  \IfFileExists{#1}{\begin{quotation}\noindent\textsc{Key:} #1\\
  \textsc{Annotation:}\ \input{#1}\end{quotation}}{}}
\providecommand{\bibAnnote}[2]{%
  \begin{quotation}\noindent\textsc{Key:} #1\\
  \textsc{Annotation:}\ #2\end{quotation}}

\bibitem[{{Asai} et~al.(2004){Asai}, {Yokoyama}, {Shimojo}, and
  {Shibata}}]{2004asaiApJL605}
{Asai}, A., {Yokoyama}, T., {Shimojo}, M., and {Shibata}, K. (2004).
\newblock {Downflow Motions Associated with Impulsive Nonthermal Emissions
  Observed in the 2002 July 23 Solar Flare}.
\newblock \emph{\apjl} 605, L77--L80.
\newblock \doi{10.1086/420768}
\bibAnnoteFile{2004asaiApJL605}

\bibitem[{{Cai} et~al.(2019){Cai}, {Shen}, {Raymond}, {Mei}, {Warmuth},
  {Roussev} et~al.}]{2019caiMNRAS489}
{Cai}, Q., {Shen}, C., {Raymond}, J.~C., {Mei}, Z., {Warmuth}, A., {Roussev},
  I.~I., et~al. (2019).
\newblock {Investigations of a supra-arcade fan and termination shock above the
  top of the flare-loop system of the 2017 September 10 event}.
\newblock \emph{\mnras} 489, 3183--3199.
\newblock \doi{10.1093/mnras/stz2167}
\bibAnnoteFile{2019caiMNRAS489}

\bibitem[{{Casini} et~al.(2017){Casini}, {White}, and
  {Judge}}]{Casinietal:2017}
{Casini}, R., {White}, S.~M., and {Judge}, P.~G. (2017).
\newblock {Magnetic Diagnostics of the Solar Corona: Synthesizing Optical and
  Radio Techniques}.
\newblock \emph{\ssr} 210, 145--181.
\newblock \doi{10.1007/s11214-017-0400-6}
\bibAnnoteFile{Casinietal:2017}

\bibitem[{{C{\'e}cere} et~al.(2012){C{\'e}cere}, {Schneiter}, {Costa},
  {Elaskar}, and {Maglione}}]{2012cecereApJ759}
{C{\'e}cere}, M., {Schneiter}, M., {Costa}, A., {Elaskar}, S., and {Maglione},
  S. (2012).
\newblock {Simulation of Descending Multiple Supra-arcade Reconnection Outflows
  in Solar Flares}.
\newblock \emph{\apj} 759, 79.
\newblock \doi{10.1088/0004-637X/759/2/79}
\bibAnnoteFile{2012cecereApJ759}

\bibitem[{{C{\'e}cere} et~al.(2015){C{\'e}cere}, {Zurbriggen}, {Costa}, and
  {Schneiter}}]{2015cecereApJ807}
{C{\'e}cere}, M., {Zurbriggen}, E., {Costa}, A., and {Schneiter}, M. (2015).
\newblock {3D MHD Simulation of Flare Supra-Arcade Downflows in a Turbulent
  Current Sheet Medium}.
\newblock \emph{\apj} 807, 6.
\newblock \doi{10.1088/0004-637X/807/1/6}
\bibAnnoteFile{2015cecereApJ807}

\bibitem[{{Chen} et~al.(2015){Chen}, {Bastian}, {Shen}, {Gary}, {Krucker}, and
  {Glesener}}]{2015chenSci350}
{Chen}, B., {Bastian}, T.~S., {Shen}, C., {Gary}, D.~E., {Krucker}, S., and
  {Glesener}, L. (2015).
\newblock {Particle acceleration by a solar flare termination shock}.
\newblock \emph{Science} 350, 1238--1242.
\newblock \doi{10.1126/science.aac8467}
\bibAnnoteFile{2015chenSci350}

\bibitem[{{Chen} et~al.(2017){Chen}, {Liu}, {Deng}, and
  {Wang}}]{2017chenA&A606}
{Chen}, X., {Liu}, R., {Deng}, N., and {Wang}, H. (2017).
\newblock {Thermodynamics of supra-arcade downflows in solar flares}.
\newblock \emph{\aap} 606, A84.
\newblock \doi{10.1051/0004-6361/201629893}
\bibAnnoteFile{2017chenA&A606}

\bibitem[{{Costa} et~al.(2009){Costa}, {Elaskar}, {Fern{\'a}ndez}, and
  {Mart{\'{\i}}nez}}]{2009costaMNRAS400}
{Costa}, A., {Elaskar}, S., {Fern{\'a}ndez}, C.~A., and {Mart{\'{\i}}nez}, G.
  (2009).
\newblock {Simulation of dark lanes in post-flare supra-arcade}.
\newblock \emph{\mnras} 400, L85--L89.
\newblock \doi{10.1111/j.1745-3933.2009.00769.x}
\bibAnnoteFile{2009costaMNRAS400}

\bibitem[{{Dulk}(1985)}]{1985dulkARAA23}
{Dulk}, G.~A. (1985).
\newblock {Radio emission from the sun and stars}.
\newblock \emph{Ann. Rev. Astron. Astrophys.} 23, 169--224
\bibAnnoteFile{1985dulkARAA23}

\bibitem[{{Freed} and {McKenzie}(2018)}]{2018freedApJ866}
{Freed}, M.~S. and {McKenzie}, D.~E. (2018).
\newblock {Quantifying Turbulent Dynamics Found within the Plasma Sheets of
  Multiple Solar Flares}.
\newblock \emph{\apj} 866, 29.
\newblock \doi{10.3847/1538-4357/aadee4}
\bibAnnoteFile{2018freedApJ866}

\bibitem[{{Gary} et~al.(2018){Gary}, {Chen}, {Dennis}, {Fleishman}, {Hurford},
  {Krucker} et~al.}]{2018garyApJ863}
{Gary}, D.~E., {Chen}, B., {Dennis}, B.~R., {Fleishman}, G.~D., {Hurford},
  G.~J., {Krucker}, S., et~al. (2018).
\newblock {Microwave and Hard X-Ray Observations of the 2017 September 10 Solar
  Limb Flare}.
\newblock \emph{\apj} 863, 83.
\newblock \doi{10.3847/1538-4357/aad0ef}
\bibAnnoteFile{2018garyApJ863}

\bibitem[{{Guo} et~al.(2014){Guo}, {Huang}, {Bhattacharjee}, and
  {Innes}}]{2014guoApJL796}
{Guo}, L.~J., {Huang}, Y., {Bhattacharjee}, A., and {Innes}, D.~E. (2014).
\newblock {Rayleigh-Taylor Type Instabilities in the Reconnection Exhaust Jet
  as a Mechanism for Supra-arcade Downflows in the Sun}.
\newblock \emph{\apjl} 796, L29.
\newblock \doi{10.1088/2041-8205/796/2/L29}
\bibAnnoteFile{2014guoApJL796}

\bibitem[{{Hanneman} and {Reeves}(2014)}]{2014hannemanApJ786}
{Hanneman}, W.~J. and {Reeves}, K.~K. (2014).
\newblock {Thermal Structure of Current Sheets and Supra-arcade Downflows in
  the Solar Corona}.
\newblock \emph{\apj} 786, 95.
\newblock \doi{10.1088/0004-637X/786/2/95}
\bibAnnoteFile{2014hannemanApJ786}

\bibitem[{Harrison and Krishnan(2012)}]{2012harrison2012python}
Harrison, C. and Krishnan, H. (2012).
\newblock {Python's Role in VisIt}.
\newblock In \emph{11th Python in Science Conf. (scipy 2012)}, eds.
  A.~{Ahmadia}, J.~{Millman}, and S.~{van der Walt}. 23--29.
\newblock \doi{10.25080/Majora-54c7f2c8-00d}
\bibAnnoteFile{2012harrison2012python}

\bibitem[{{Hayes} et~al.(2019){Hayes}, {Gallagher}, {Dennis}, {Ireland},
  {Inglis}, and {Morosan}}]{2019hayesApJ875}
{Hayes}, L.~A., {Gallagher}, P.~T., {Dennis}, B.~R., {Ireland}, J., {Inglis},
  A., and {Morosan}, D.~E. (2019).
\newblock {Persistent Quasi-periodic Pulsations during a Large X-class Solar
  Flare}.
\newblock \emph{\apj} 875, 33.
\newblock \doi{10.3847/1538-4357/ab0ca3}
\bibAnnoteFile{2019hayesApJ875}

\bibitem[{{Innes} et~al.(2014){Innes}, {Guo}, {Bhattacharjee}, {Huang}, and
  {Schmit}}]{2014innesApJ796}
{Innes}, D.~E., {Guo}, L.~J., {Bhattacharjee}, A., {Huang}, Y.~M., and
  {Schmit}, D. (2014).
\newblock {Observations of Supra-arcade Fans: Instabilities at the Head of
  Reconnection Jets}.
\newblock \emph{\apj} 796, 27.
\newblock \doi{10.1088/0004-637X/796/1/27}
\bibAnnoteFile{2014innesApJ796}

\bibitem[{{Innes} et~al.(2003{\natexlab{a}}){Innes}, {McKenzie}, and
  {Wang}}]{2003innesSoPh217.267}
{Innes}, D.~E., {McKenzie}, D.~E., and {Wang}, T. (2003{\natexlab{a}}).
\newblock {Observations of 1000 km s$^{-1}$ Doppler shifts in 10$^{7}$ K solar
  flare supra-arcade}.
\newblock \emph{\solphys} 217, 267--279
\bibAnnoteFile{2003innesSoPh217.267}

\bibitem[{{Innes} et~al.(2003{\natexlab{b}}){Innes}, {McKenzie}, and
  {Wang}}]{2003innesSoPh217.247}
{Innes}, D.~E., {McKenzie}, D.~E., and {Wang}, T. (2003{\natexlab{b}}).
\newblock {SUMER spectral observations of post-flare supra-arcade inflows}.
\newblock \emph{\solphys} 217, 247--265
\bibAnnoteFile{2003innesSoPh217.247}

\bibitem[{Kerdraon and Delouis(1997)}]{KerdraonDelouis:1997}
Kerdraon, A. and Delouis, J.-M. (1997).
\newblock The nan{\c{c}}ay radioheliograph.
\newblock In \emph{Coronal Physics from Radio and Space Observations}, ed.
  G.~Trottet (Springer Berlin Heidelberg), 192--201.
\newblock \doi{10.1007/BFb0106458}
\bibAnnoteFile{KerdraonDelouis:1997}

\bibitem[{{Li} et~al.(2021){Li}, {Cheng}, {Ding}, {Reeves}, {Kittrell}, {Weber}
  et~al.}]{2021liApJ915}
{Li}, Z.~F., {Cheng}, X., {Ding}, M.~D., {Reeves}, K.~K., {Kittrell}, D.,
  {Weber}, M., et~al. (2021).
\newblock {Thermodynamic Evolution of Solar Flare Supra-arcade Downflows}.
\newblock \emph{\apj} 915, 124.
\newblock \doi{10.3847/1538-4357/ac043e}
\bibAnnoteFile{2021liApJ915}

\bibitem[{{Linton} et~al.(2009){Linton}, {Devore}, and
  {Longcope}}]{2009lintonEPS61}
{Linton}, M.~G., {Devore}, C.~R., and {Longcope}, D.~W. (2009).
\newblock {Patchy reconnection in a Y-type current sheet}.
\newblock \emph{Earth, Planets, and Space} 61, 573--576
\bibAnnoteFile{2009lintonEPS61}

\bibitem[{{Linton} and {Longcope}(2006)}]{2006lintonApJ642}
{Linton}, M.~G. and {Longcope}, D.~W. (2006).
\newblock {A Model for Patchy Reconnection in Three Dimensions}.
\newblock \emph{\apj} 642, 1177--1192.
\newblock \doi{10.1086/500965}
\bibAnnoteFile{2006lintonApJ642}

\bibitem[{{Liu}(2013)}]{2013liuMNRAS434}
{Liu}, R. (2013).
\newblock {Dynamical processes at the vertical current sheet behind an erupting
  flux rope}.
\newblock \emph{\mnras} 434, 1309--1320.
\newblock \doi{10.1093/mnras/stt1090}
\bibAnnoteFile{2013liuMNRAS434}

\bibitem[{{Liu} et~al.(2013){Liu}, {Chen}, and {Petrosian}}]{2013liuApJ767}
{Liu}, W., {Chen}, Q., and {Petrosian}, V. (2013).
\newblock {Plasmoid Ejections and Loop Contractions in an Eruptive M7.7 Solar
  Flare: Evidence of Particle Acceleration and Heating in Magnetic Reconnection
  Outflows}.
\newblock \emph{\apj} 767, 168.
\newblock \doi{10.1088/0004-637X/767/2/168}
\bibAnnoteFile{2013liuApJ767}

\bibitem[{{Longcope} et~al.(2018){Longcope}, {Unverferth}, {Klein}, {McCarthy},
  and {Priest}}]{2018longcopeApJ868}
{Longcope}, D., {Unverferth}, J., {Klein}, C., {McCarthy}, M., and {Priest}, E.
  (2018).
\newblock {Evidence for Downflows in the Narrow Plasma Sheet of 2017 September
  10 and Their Significance for Flare Reconnection}.
\newblock \emph{\apj} 868, 148.
\newblock \doi{10.3847/1538-4357/aaeac4}
\bibAnnoteFile{2018longcopeApJ868}

\bibitem[{{Longcope} et~al.(2009){Longcope}, {Guidoni}, and
  {Linton}}]{2009longcopeApJL690}
{Longcope}, D.~W., {Guidoni}, S.~E., and {Linton}, M.~G. (2009).
\newblock {Gas-dynamic Shock Heating of Post-flare Loops Due to Retraction
  Following Localized, Impulsive Reconnection}.
\newblock \emph{\apjl} 690, L18--L22.
\newblock \doi{10.1088/0004-637X/690/1/L18}
\bibAnnoteFile{2009longcopeApJL690}

\bibitem[{{Maglione} et~al.(2011){Maglione}, {Schneiter}, {Costa}, and
  {Elaskar}}]{2011maglioneAA527}
{Maglione}, L.~S., {Schneiter}, E.~M., {Costa}, A., and {Elaskar}, S. (2011).
\newblock {Simulation of dark lanes in post-flare supra-arcades. III. A 2D
  simulation}.
\newblock \emph{\aap} 527, L5.
\newblock \doi{10.1051/0004-6361/201015934}
\bibAnnoteFile{2011maglioneAA527}

\bibitem[{{McKenzie}(2013)}]{2013mackenzieApJ766}
{McKenzie}, D.~E. (2013).
\newblock {Turbulent Dynamics in Solar Flare Sheet Structures Measured with
  Local Correlation Tracking}.
\newblock \emph{\apj} 766, 39.
\newblock \doi{10.1088/0004-637X/766/1/39}
\bibAnnoteFile{2013mackenzieApJ766}

\bibitem[{{McKenzie} and {Hudson}(1999)}]{1999mckenzieApJL519}
{McKenzie}, D.~E. and {Hudson}, H.~S. (1999).
\newblock {X-Ray Observations of Motions and Structure above a Solar Flare
  Arcade}.
\newblock \emph{\apjl} 519, L93--L96.
\newblock \doi{10.1086/312110}
\bibAnnoteFile{1999mckenzieApJL519}

\bibitem[{{McKenzie} and {Savage}(2009)}]{2009mckenzieApJ697}
{McKenzie}, D.~E. and {Savage}, S.~L. (2009).
\newblock {Quantitative Examination of Supra-arcade Downflows in Eruptive Solar
  Flares}.
\newblock \emph{\apj} 697, 1569--1577.
\newblock \doi{10.1088/0004-637X/697/2/1569}
\bibAnnoteFile{2009mckenzieApJ697}

\bibitem[{{McKenzie} and {Savage}(2011)}]{2011mackenzieApJL735}
{McKenzie}, D.~E. and {Savage}, S.~L. (2011).
\newblock {Distribution Functions of Sizes and Fluxes Determined from
  Supra-arcade Downflows}.
\newblock \emph{\apjl} 735, L6.
\newblock \doi{10.1088/2041-8205/735/1/L6}
\bibAnnoteFile{2011mackenzieApJL735}

\bibitem[{{Prialnik}(2000)}]{2000prialnikBook}
{Prialnik}, D. (2000).
\newblock \emph{{``An Introduction to the Theory of Stellar Structure and
  Evolution''}} (Cambridge: University Press, 2000).
\newblock \doi{10.1080/00107514.2011.580371}
\bibAnnoteFile{2000prialnikBook}

\bibitem[{{Reeves} et~al.(2017){Reeves}, {Freed}, {McKenzie}, and
  {Savage}}]{2017reevesApJ836}
{Reeves}, K.~K., {Freed}, M.~S., {McKenzie}, D.~E., and {Savage}, S.~L. (2017).
\newblock {An Exploration of Heating Mechanisms in a Supra-arcade Plasma Sheet
  Formed after a Coronal Mass Ejection}.
\newblock \emph{\apj} 836, 55.
\newblock \doi{10.3847/1538-4357/836/1/55}
\bibAnnoteFile{2017reevesApJ836}

\bibitem[{{Savage} and {McKenzie}(2011)}]{2011savageApJ730}
{Savage}, S.~L. and {McKenzie}, D.~E. (2011).
\newblock {Quantitative Examination of a Large Sample of Supra-arcade Downflows
  in Eruptive Solar Flares}.
\newblock \emph{\apj} 730, 98.
\newblock \doi{10.1088/0004-637X/730/2/98}
\bibAnnoteFile{2011savageApJ730}

\bibitem[{{Savage} et~al.(2012){Savage}, {McKenzie}, and
  {Reeves}}]{2012savageApJL747}
{Savage}, S.~L., {McKenzie}, D.~E., and {Reeves}, K.~K. (2012).
\newblock {Re-interpretation of Supra-arcade Downflows in Solar Flares}.
\newblock \emph{\apjl} 747, L40.
\newblock \doi{10.1088/2041-8205/747/2/L40}
\bibAnnoteFile{2012savageApJL747}

\bibitem[{{Schulz} et~al.(2010){Schulz}, {Costa}, {Elaskar}, and
  {Cid}}]{2010schultzMNRAS407}
{Schulz}, W., {Costa}, A., {Elaskar}, S., and {Cid}, G. (2010).
\newblock {Simulation of dark lanes in post-flare supra-arcades - II. A
  contribution to the remote sensing of the coronal magnetic field}.
\newblock \emph{\mnras} 407, L89--L93.
\newblock \doi{10.1111/j.1745-3933.2010.00911.x}
\bibAnnoteFile{2010schultzMNRAS407}

\bibitem[{{Scott} et~al.(2013){Scott}, {Longcope}, and
  {McKenzie}}]{2013scottApJ776}
{Scott}, R.~B., {Longcope}, D.~W., and {McKenzie}, D.~E. (2013).
\newblock {Peristaltic Pumping near Post-coronal Mass Ejection Supra-arcade
  Current Sheets}.
\newblock \emph{\apj} 776, 54.
\newblock \doi{10.1088/0004-637X/776/1/54}
\bibAnnoteFile{2013scottApJ776}

\bibitem[{{Selhorst} et~al.(2008){Selhorst}, {Silva-V{\'a}lio}, and
  {Costa}}]{Selhorstetal:2008}
{Selhorst}, C.~L., {Silva-V{\'a}lio}, A., and {Costa}, J.~E.~R. (2008).
\newblock {Solar atmospheric model over a highly polarized 17 GHz active
  region}.
\newblock \emph{\aap} 488, 1079--1084.
\newblock \doi{10.1051/0004-6361:20079217}
\bibAnnoteFile{Selhorstetal:2008}

\bibitem[{{van Haarlem} et~al.(2013){van Haarlem}, {Wise}, {Gunst}, {Heald},
  {McKean}, {Hessels} et~al.}]{vanHaarlemetal:2013}
{van Haarlem}, M.~P., {Wise}, M.~W., {Gunst}, A.~W., {Heald}, G., {McKean},
  J.~P., {Hessels}, J.~W.~T., et~al. (2013).
\newblock {LOFAR: The LOw-Frequency ARray}.
\newblock \emph{\aap} 556, A2.
\newblock \doi{10.1051/0004-6361/201220873}
\bibAnnoteFile{vanHaarlemetal:2013}

\bibitem[{{van Hoof} et~al.(2014){van Hoof}, {Williams}, {Volk}, {Chatzikos},
  {Ferland}, {Lykins} et~al.}]{vanHoofetal:2014}
{van Hoof}, P.~A.~M., {Williams}, R.~J.~R., {Volk}, K., {Chatzikos}, M.,
  {Ferland}, G.~J., {Lykins}, M., et~al. (2014).
\newblock {Accurate determination of the free-free Gaunt factor - I.
  Non-relativistic Gaunt factors}.
\newblock \emph{\mnras} 444, 420--428.
\newblock \doi{10.1093/mnras/stu1438}
\bibAnnoteFile{vanHoofetal:2014}

\bibitem[{{Verwichte} et~al.(2005){Verwichte}, {Nakariakov}, and
  {Cooper}}]{2005verwichteA&A430}
{Verwichte}, E., {Nakariakov}, V.~M., and {Cooper}, F.~C. (2005).
\newblock {Transverse waves in a post-flare supra-arcade}.
\newblock \emph{\aap} 430, L65--L68.
\newblock \doi{10.1051/0004-6361:200400133}
\bibAnnoteFile{2005verwichteA&A430}

\bibitem[{{Warren} et~al.(2018){Warren}, {Brooks}, {Ugarte-Urra}, {Reep},
  {Crump}, and {Doschek}}]{2018warrenApJ854}
{Warren}, H.~P., {Brooks}, D.~H., {Ugarte-Urra}, I., {Reep}, J.~W., {Crump},
  N.~A., and {Doschek}, G.~A. (2018).
\newblock {Spectroscopic Observations of Current Sheet Formation and
  Evolution}.
\newblock \emph{\apj} 854, 122.
\newblock \doi{10.3847/1538-4357/aaa9b8}
\bibAnnoteFile{2018warrenApJ854}

\bibitem[{{Warren} et~al.(2011){Warren}, {O'Brien}, and
  {Sheeley}}]{2011warrenApJ742}
{Warren}, H.~P., {O'Brien}, C.~M., and {Sheeley}, J., Neil~R. (2011).
\newblock {Observations of Reconnecting Flare Loops with the Atmospheric
  Imaging Assembly}.
\newblock \emph{\apj} 742, 92.
\newblock \doi{10.1088/0004-637X/742/2/92}
\bibAnnoteFile{2011warrenApJ742}

\bibitem[{{White}(2004)}]{White:2004}
{White}, S.~M. (2004).
\newblock {Coronal Magnetic Field Measurements Through Gyroresonance Emission}.
\newblock In \emph{Astrophysics and Space Science Library}, eds. D.~E. {Gary}
  and C.~U. {Keller}. vol. 314 of \emph{Astrophysics and Space Science
  Library}, 89.
\newblock \doi{10.1007/1-4020-2814-8\_5}
\bibAnnoteFile{White:2004}

\bibitem[{{Xue} et~al.(2020){Xue}, {Su}, {Li}, and {Zhao}}]{2020xueApJ898}
{Xue}, J., {Su}, Y., {Li}, H., and {Zhao}, X. (2020).
\newblock {Thermodynamical Evolution of Supra-arcade Downflows}.
\newblock \emph{\apj} 898, 88.
\newblock \doi{10.3847/1538-4357/ab9a3d}
\bibAnnoteFile{2020xueApJ898}

\bibitem[{{Yu} et~al.(2020){Yu}, {Chen}, {Reeves}, {Gary}, {Musset},
  {Fleishman} et~al.}]{2020yuApJ900}
{Yu}, S., {Chen}, B., {Reeves}, K.~K., {Gary}, D.~E., {Musset}, S.,
  {Fleishman}, G.~D., et~al. (2020).
\newblock {Magnetic Reconnection during the Post-impulsive Phase of a
  Long-duration Solar Flare: Bidirectional Outflows as a Cause of Microwave and
  X-Ray Bursts}.
\newblock \emph{\apj} 900, 17.
\newblock \doi{10.3847/1538-4357/aba8a6}
\bibAnnoteFile{2020yuApJ900}

\bibitem[{{Zurbriggen} et~al.(2016){Zurbriggen}, {Costa}, {Esquivel},
  {Schneiter}, and {C{\'e}cere}}]{2016zurbriggenApJ832}
{Zurbriggen}, E., {Costa}, A., {Esquivel}, A., {Schneiter}, M., and
  {C{\'e}cere}, M. (2016).
\newblock {MHD Simulations of Coronal Supra-arcade Downflows Including
  Anisotropic Thermal Conduction}.
\newblock \emph{\apj} 832, 74.
\newblock \doi{10.3847/0004-637X/832/1/74}
\bibAnnoteFile{2016zurbriggenApJ832}

\end{thebibliography}

\end{document}